\documentclass[review]{elsarticle}

\usepackage{lineno,hyperref}
\usepackage{subcaption}
\usepackage{graphicx}
\usepackage{amsmath}
\usepackage{array}
\usepackage{float}
\usepackage{siunitx}
\usepackage{setspace}
\modulolinenumbers[5]

\journal{International Journal of Heat and Fluid Flow}

\begin{document}

\begin{frontmatter}

\title{Application of a POD-Galerkin based method to time resolved and time unresolved data for the determination of the Convective Velocity of Large-Scale Coherent Structures in High Speed Flows}

\author{T. Sikroria}
\address{Department of Mechanical Engineering, University of Melbourne, Parkville Campus, Melbourne, Victoria - 3010, Australia}

\cortext[mycorrespondingauthor]{Corresponding author}
\ead{tsikroria@student.unimelb.edu.au}


\author{J. Soria}
\address{Laboratory for Turbulence Research in Aerospace and Combustion (LTRAC), Department of Mechanical and Aerospace Engineering, Monash University, Clayton Campus, Melbourne, Victoria - 3800, Australia}

\author{R. Sandberg, A. Ooi}
\address{Department of Mechanical Engineering, University of Melbourne, Parkville Campus, Melbourne, Victoria - 3010, Australia}

\begin{abstract}
Motivated by the aero-acoustic feedback loop phenomenon in high speed free jets and impinging jets, a thorough examination of a POD (Proper Orthogonal Decomposition)-Galerkin method to determine the average convection velocity of coherent structures in the shear layer is presented in this paper. The technique is shown to be applicable to both time resolved as well as time unresolved data, if the data set meets certain requirements. Using a detailed sensitivity analysis on a synthetic data set, a quantitative estimate on the required time resolution for the technique has been found, which can be useful for both experimental, as well as numerical studies investigating the aero-acoustic feedback loop in high speed flows. Moreover, some innovative ways to apply the technique are also demonstrated using a simulated data set, showing the effectiveness of the technique to any general problem in supersonic jets, heat transfer, combustion or other areas in fluid mechanics, where an advection process can be identified.  
\end{abstract}

\begin{keyword}
high speed jets \sep convection velocity \sep POD \sep advection model
\MSC[2019] 
\end{keyword}

\end{frontmatter}

\doublespacing

\section{Introduction}
Practical high speed jets in the transonic or supersonic flow regime, often lead to distinct resonant tones (screech tones in supersonic free jets and impingement tones in high speed impinging jets) due to the aero-acoustic feedback loop phenomena. An important feature of the flow dynamics of such turbulent flows is the shear layer instability triggered by the upstream propagating acoustic waves at the phase locked frequency. These instabilities lead to the generation of large-scale coherent vortical structures in the shear layer which are convected downstream at high subsonic or supersonic speed.  Understanding the shear layer instabilities in such flows is vital for applications like gas turbine engine exhaust noise, acoustics in impinging jets (VTOL aircraft) and surface finish in cold gas spray additive manufacturing processes. 

Brown and Roshko \cite{Brown_Roshko} have shown that large-scale structures travel downstream at a nearly constant speed with the size and spacing between them increasing as they convect downstream. As discussed by \cite{Tam_Review,Krothapalli}, the convection velocity of the large scale periodic structures in the shear layer is one of the important quantifiable parameters for the investigation of the flow instabilities. A number of methods to estimate the convection velocity in turbulent free shear flows have been used in the past and are summarized by \cite{Literature}. The most widely used technique in turbulent flows is based on the two-point correlation. As discussed by \cite{Goldschmidt}, the methodology involves the cross-correlation of time resolved data sets at two different spatial locations in the flow field and obtaining the time shift ($\Delta \tau$) which gives the maximum correlation. If the spatial separation between the two points is represented by $\Delta x$, the convection velocity can be estimated to be
\begin{equation}
U_c \approx \frac{\Delta x}{\Delta \tau} .
\end{equation}
This approach requires high time resolution and is computationally expensive if the convection velocity ($U_c$) is to be determined at multiple locations in a spatial domain. While the technique is useful for low speed turbulent flows where the data is obtained using hot-wire measurements, the application has limitations on the data obtained using imaging-based techniques like particle image velocimetry (PIV) due to its limited spatial and temporal resolutions. A phase-velocity based approach was proposed by \cite{Ganapathi} for time resolved PIV data, where the convection velocity per wavenumber is determined using the phase difference over a given time period. If the time resolution of the data is denoted by $\Delta t$ and the angle of the spatial cross-spectrum between the velocity field at $t$ and $t + \Delta t$, by $\psi (k_{\mathbf{x}},t)$, the convection velocity is calculated using
\begin{equation}
U_c (k_{\mathbf{x}},t) = \frac{\psi (k_{\mathbf{x}},t)}{2 \pi k_{\mathbf{x}} \Delta t} .
\end{equation}
However, due to the limited spatial resolution of PIV data, the high frequency structures are filtered out, which is not desirable in the research applications involving an aero-acoustic feedback loop. An improved modification of the technique was proposed by \cite{renard2015scale} where the frequency dependent convection velocity was computed using spatial derivatives and finite span time signals. However, the method has limitations in the applications where the validity of the Taylor's hypothesis is limited as the group velocity is not defined from the exact wave equation describing the problem.

The computation of the convection velocity in high speed flows is somewhat more challenging, as the data needs to have even better time resolution and a large number of samples. Time resolved experimental measurements using PIV and Schlieren's techniques require very high temporal resolution cameras. While time resolution might not be an issue for  numerical analysis using direct numerical simulations (DNS), a long time series of data needs to be stored, requiring excessive computational resources. With advancement in technology, ultra high speed optical measurements at an acquisition rate of one MHz or higher have been used in the past decade to capture supersonic flow features using Schlieren's and shadowgraph techniques (\cite{HS_Imaging_1, HS_Imaging_2, HS_Imaging_3}). Blohm et al. \cite{MHz_measurement} have applied the space-time correlation method on time resolved images to obtain the convection velocity. Murray and Lyons \cite{Mach_wave} have estimated the convection velocity by tracing the Mach wave angles in the high speed images. However, as discussed by \cite{Literature}, the drawback of these methodologies lies in the fact that the convection velocity of the local flow field is obtained rather than the exclusive behavior of the large-scale coherent structures. As the large-scale structures travel with different convection speed than the smaller scales, the estimation using the above methods is likely to be inaccurate.  

As defined by Hussain \cite{Hussain}, a coherent structure is a connected turbulent fluid mass with instantaneously phase-correlated vorticity over its spatial extent. However, time-resolved measurements of vorticity over a two-dimensional domain in high-speed flow is difficult. Considering the fact that the data is available in the form of images, Thurow et al. \cite{Literature} have loosely defined a large-scale structure as any feature of the flow that has a size which roughly spans the observable shear layer. Krothapalli et al. \cite{Krothapalli} have used time unresolved PIV data to trace such vortex structures. For each of these identified structures, the velocity value at the location having the maximum vorticity was deemed to represent the instantaneous convection velocity of that structure. An average value of the convection velocity, obtained using ensemble averaging, was reported at various spatial locations. While the method considers only the large-scale vortices, it is computationally expensive. Adelgren et al. \cite{Jet_control} have applied physical control mechanisms to generate stable vortex structures, followed by imaging. The convection velocity was determined using the same approach as \cite{Krothapalli}. However, the method is application specific and cannot be applied to a general problem involving aero-acoustic feedback loop in supersonic jets. Weightman et al. \cite{Joel} have used a phase averaging method for supersonic impinging jets. This technique however, requires the additional input of the lock-in frequency which needs to be obtained using microphone measurements.   

A simpler approach has been proposed by Jaunet et al. \cite{Vincent_Jaunet} using a POD-Galerkin advection model to estimate the average convection velocity in a domain. The technique is applicable to any kind of data which meets certain conditions specific to convective flow. The conditions have been discussed by Jaunet et al. \cite{Vincent_Jaunet} and also illustrated in the following section. The question arises if the technique can be applied to a time unresolved data and also if the required time resolution can be quantified in some way in the case of time resolved data. These aspects have been addressed in this manuscript.

\section{Methodology} \label{s_method}

The method is based on the proper orthogonal decomposition (POD) of a function. The theory and details of POD can be found in \cite{Chatterjee}, \cite{lumley1967structure}, \cite{berkooz1993proper} and \cite{holmes2012turbulence} .Let $f$ be a general function of space and time which is decomposed using POD into some finite $K$ modes as
\begin{equation} \label{e_POD}
f(\mathbf{x},t) \simeq \sum^K_{k=1} a_k (t) \mathbf{\phi_k} (\mathbf{x}). 
\end{equation}
As proposed by \cite{Vincent_Jaunet}, if the function is purely advective, it will satisfy the advection equation,
\begin{equation} \label{e_adv}
\frac{\partial f }{\partial t} + \mathbf{U_c \cdot \nabla}f = 0 \: \: \mbox{or} \:  \: \frac{\partial f }{\partial t} + U_c \frac{\partial f }{\partial x} = 0 \: \mbox{(for \: 1-D)}.
\end{equation}
Substituting $f$, as approximated by equation \ref{e_POD}, into equation \ref{e_adv} results in
\begin{equation} \label{e_POD_adv}
\sum^K_{k=1}=1 \frac{\partial a_k}{\partial t} \phi_k + U_c \sum^K_{k=1} a_k \frac{\partial \phi_k}{\partial x} = 0 \: \mbox{or} \: \sum^K_{k=1}=1 a_{k,t} \phi_k + U_c \sum^K_{k=1} a_k \phi_{k,x} = 0.
\end{equation}
Take the inner product of equation \ref{e_POD_adv} with $\phi_m$ and note that the modes ($\mathbf{\phi_k}$) are orthonormal, i.e.  
\begin{equation}
(\phi_k, \phi_m) = \delta_{km} ,
\end{equation} 
and the coefficients ($a_k$) are orthogonal, i.e.
\begin{equation}
<a_k a_m> = \delta_{km} \lambda_m . 
\end{equation}
The symbol '$<>$' represents ensemble average (as the coefficients are only a function of time) and the expression '$(,)$' represents the inner product. 

Furthermore, taking the projection of a general $p^{th}$ mode on the advection equation yields 
\begin{equation}
\sum^K_{k=1} a_{k,t} (\phi_k , \phi_p ) + U_c \sum^K_{k=1} a_k ( \phi_{k,x}, \phi_p ) = 0.
\end{equation}
Given the orthonormal property of the spatial modes, this yields 
\begin{equation}
a_{p,t} + U_c \sum^K_{k=1} a_k ( \phi_{k,x}, \phi_p ) = 0.
\end{equation}
Multiplying the resultant equation by a general $n^{th}$ mode followed by ensemble averaging results in  
\begin{equation}
<a_{p,t} a_n > + \: U_c < a_n \sum^K_{k=1} a_k ( \phi_{k,x}, \phi_p ) > = 0.
\end{equation}
The summation is over the spatial modes and ensemble averaging is over time. As the two are independent, the ensemble operator is brought inside summation, leading to
\begin{equation}
<a_{p,t} a_n > + \: U_c \sum^K_{k=1} <a_n a_k> ( \phi_{k,x}, \phi_p ) = 0.
\end{equation}  
Using the orthogonal properties of the temporal coefficients yields
\begin{equation} \label{e_final}
<a_{p,t} a_n > + \: U_c <a_n a_n > ( \phi_{n,x}, \phi_p ) = 0. 
\end{equation}
If $p=n$,
\begin{equation}
<a_{n,t} a_n > = \frac{1}{2} \frac{\partial <a_n^2>}{\partial t} = \frac{1}{2} \frac{\partial \lambda_n}{\partial t} = 0.
\end{equation}
This result when applied to equation \ref{e_final} leads to the test condition for advection given by
\begin{equation} \label{e_test_con_or}
(\phi_{n,x} \phi_n) = 0.
\end{equation}
If $p\not = n$, we get the expression for convection velocity as
\begin{equation} \label{e_conv}
U_c = \frac{- <a_{p,t} a_n >}{<a_n^2 > ( \phi_{n,x}, \phi_p )}. 
\end{equation}
For a convective flow having non-zero convection velocity, the numerator in equation \ref{e_conv} will not be zero. It can only be zero if there is only a single dominant mode rather than a mode pair which implies zero convection velocity or a standing wave.

For discrete data, the coefficients will be available as data values in the time domain (discretized into \textit{N} points). Hence, the $a_k$'s will be column vectors of size $N$ which are orthogonal to each other. The inner product of a column vector with itself will give the eigenvalue of the mode ($(a_n,a_n) = \lambda_n$). The ensemble average for discrete data is 
$$ <a_n a_n > = \frac{\sum^{N_n}_{i=1} a_{ni}^2 \Delta T}{T} = \frac{(a_n,a_n)}{N_{a_n}} ,$$
where $N$ represents the number of samples in the corresponding discrete data set. Rewriting equation \ref{e_conv} for a discrete data set yields
\begin{equation} \label{e_gfinal}
U_c = \frac{- <a_{p,t} a_n >}{<a_n^2 > ( \phi_{n,x}, \phi_p )} = \frac{- \frac{(a_{p,t} a_n )}{N_{a_p}}}{ \frac{\lambda_n}{N_{a_n}} ( \phi_{n,x}, \phi_p )} .
\end{equation}
As evident from equation \ref{e_gfinal}, the convection velocity can be determined using the instantaneous temporal derivatives of the coefficients and the spatial derivatives of the modes. The accuracy of prediction will have some sensitivity to both the spatial and the temporal resolution of the data set. Moreover, the proposed methodology is applicable only to non-dispersive waves as it is based on a constant convection velocity, shown in equation \ref{e_adv}.

For practical data (computational or experimental), the derivatives have to be evaluated using finite difference stencils. The inner products are also computed numerically using matrix multiplication of one vector with the transpose of the other. Depending on the spatial grid resolution ($\Delta x$) as well as the discretization schemes, there will be some numerical errors in the evaluation of the above expressions. The test condition (\ref{e_test_con_or}) will thus, not give exactly zero but some small number. It is therefore, prudent to express the test condition for a discrete data in non-dimensional form, specified as 
\begin{equation} \label{e_test_con}
(\phi_{n,x} \phi_n) \Delta x = \epsilon , \: \epsilon \rightarrow 0 .
\end{equation} 
There are no defined criteria for $\epsilon$ and the choice depends on the type of data and the problem. The best approach is to plot the values of $\epsilon$ for the first few mode combinations and consider their the relative values. The two modes representing advection will have a very low value of $\epsilon$ relative to the other modes. These modes are selected for the application of the technique. As illustrated in \cite{Vincent_Jaunet}, the contour plot of the inner product on a $p-n$ domain, clearly shows very low values at $n=p$ for the convective modes. For practical applications, such pair of convective modes are considered which satisfy the test conditions. The most dominant mode pair is then considered to estimate the convection velocity of the large scale structures. In case the flow dynamics in some research applications (an example of such application is shown in \cite{Joel_JFM_2}) show multiple dominant mode pairs having comparable energy, the methodology can be applied to each pair individually, giving its respective convection velocity.

\section{Application to high speed flows}
\begin{figure}
	\centerline{\includegraphics[scale=0.325]{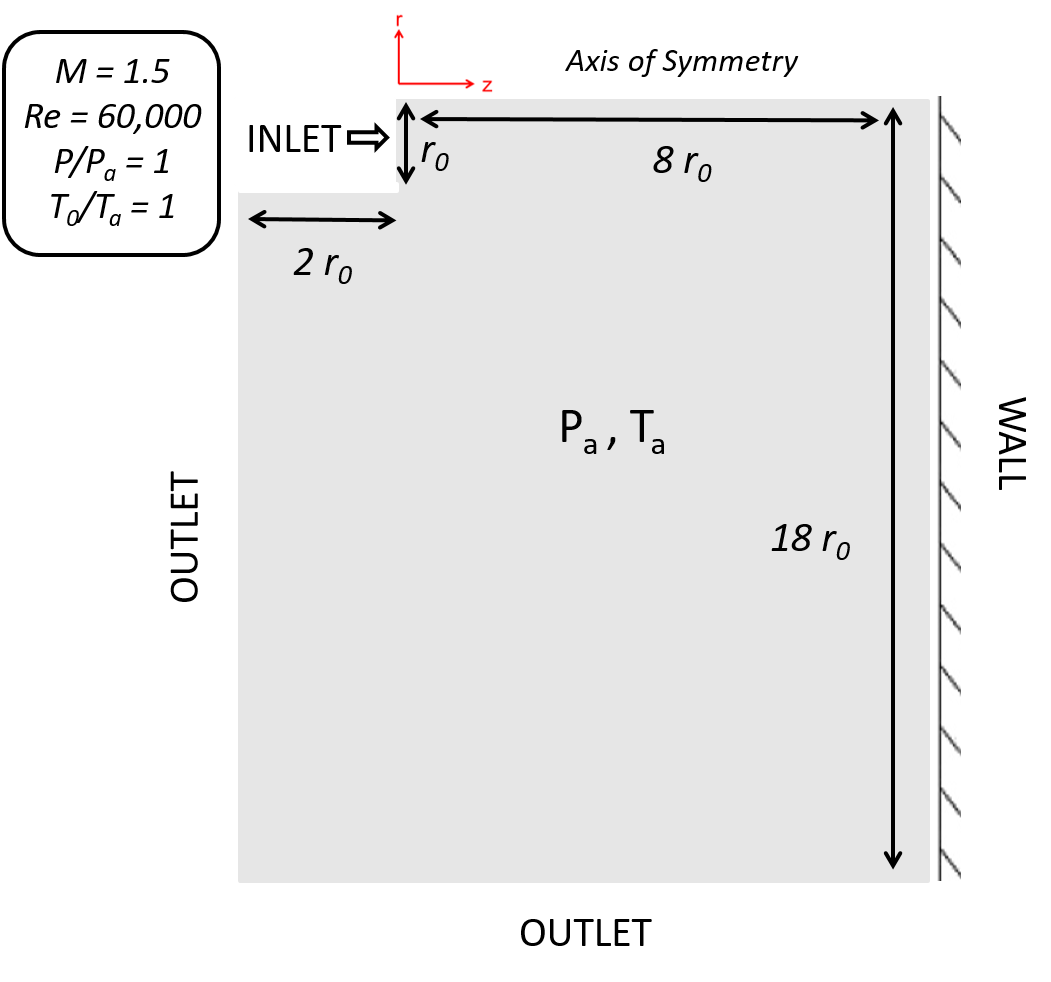}}
	\caption{Simulation Domain and Conditions}
	\label{f_Simulation}
\end{figure}

In order to demonstrate the application of the proposed methodology, a simulated flow of an ideally expanded supersonic impinging jet at Mach 1.5 has been used as the test data set. An axi-symmetric simulation was carried out using OpenFOAM with the same domain and operating conditions as those used by \cite{gojon}. The details are shown in figure \ref{f_Simulation}. The simulation mimics one of the experiments by \cite{Krothapalli}, except for the fact that the Reynolds number for the simulation is an order of magnitude lower. The mesh has been kept identical to that used by \cite{gojon}, i.e. a spatial grid of 400 x 600 has been used, with finer mesh in the shear layer and wall-jet regions and coarser mesh towards the outlets. Following the approach of \cite{gojon}, the computational time step is chosen as
\begin{equation}
\delta t = 0.005 \frac{r_0}{U_j} .
\end{equation}
$U_j$ refers to the ideally expanded jet velocity at the inlet. Defining a flow through time as
\begin{equation}
t_f = 8 \frac{r_0}{U_j} ,
\end{equation}
which gives an estimate of the time taken by a fluid parcel along the centerline to reach the impinging wall, a time scale of 1600 computational time steps represents one flow through time. Neglecting the initial 6 flow though time scales as they are considered to be transient, the data is saved after every 100 time steps, which is hence, the time resolution of the acquired data, i.e. 
\begin{equation}
\Delta t = 0.5 \frac{r_0}{U_j} = \frac{t_f}{16} .
\end{equation}
One flow through time scale is thus, resolved into 16 snapshots. The solver used for the simulations was \textit{rhoEnergyFOAM}, which has high accuracy but takes a long time for computations. Hence, due to computational constraints, only 80 samples of data were used for the analysis, representing a time span of 5 times the flow through time.

\begin{figure}
	\centerline{\includegraphics[scale=0.18]{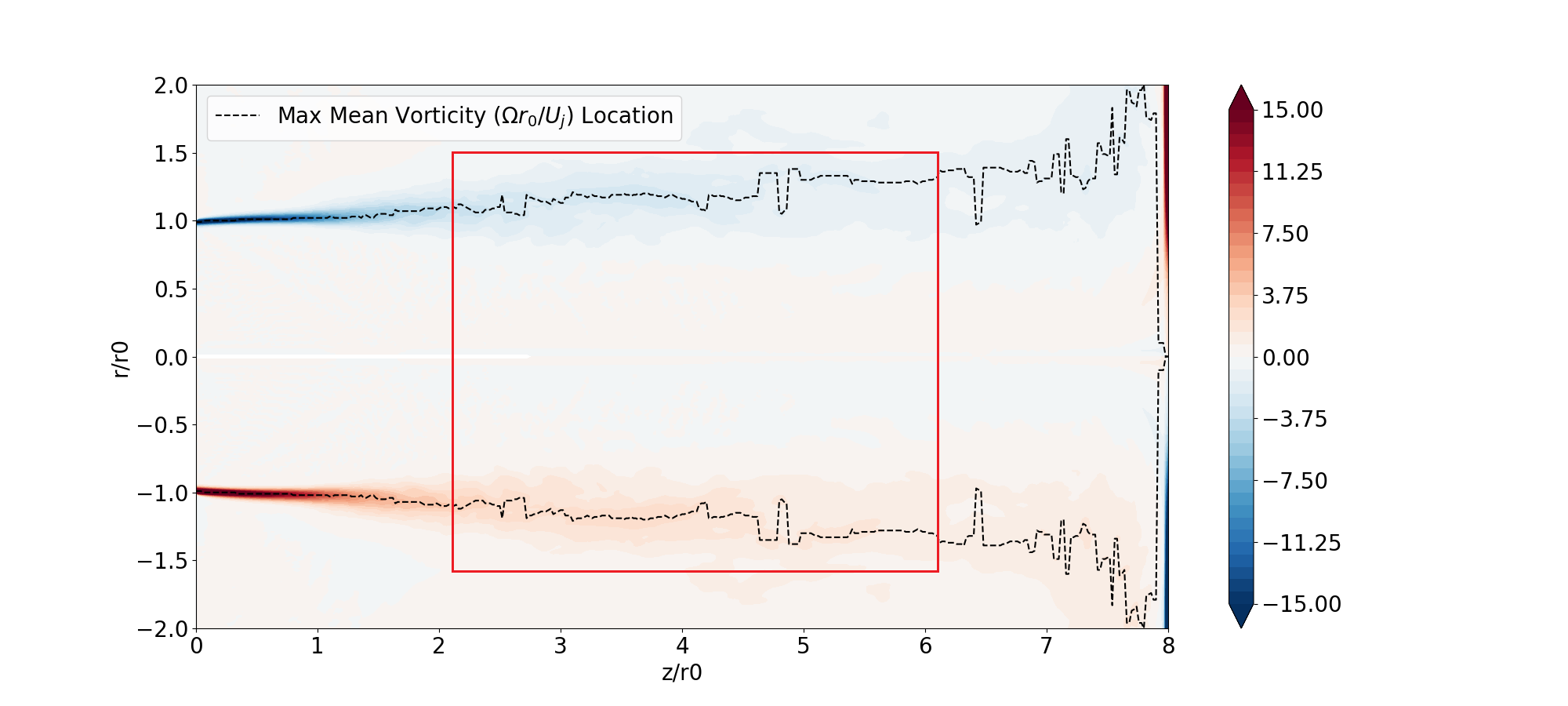}}
	\caption{Mean Vorticity Contours with the location of maximum vorticity}
	\label{f_Mean_Vorticity}
\end{figure}

\begin{figure}
	\centerline{\includegraphics[scale=0.3]{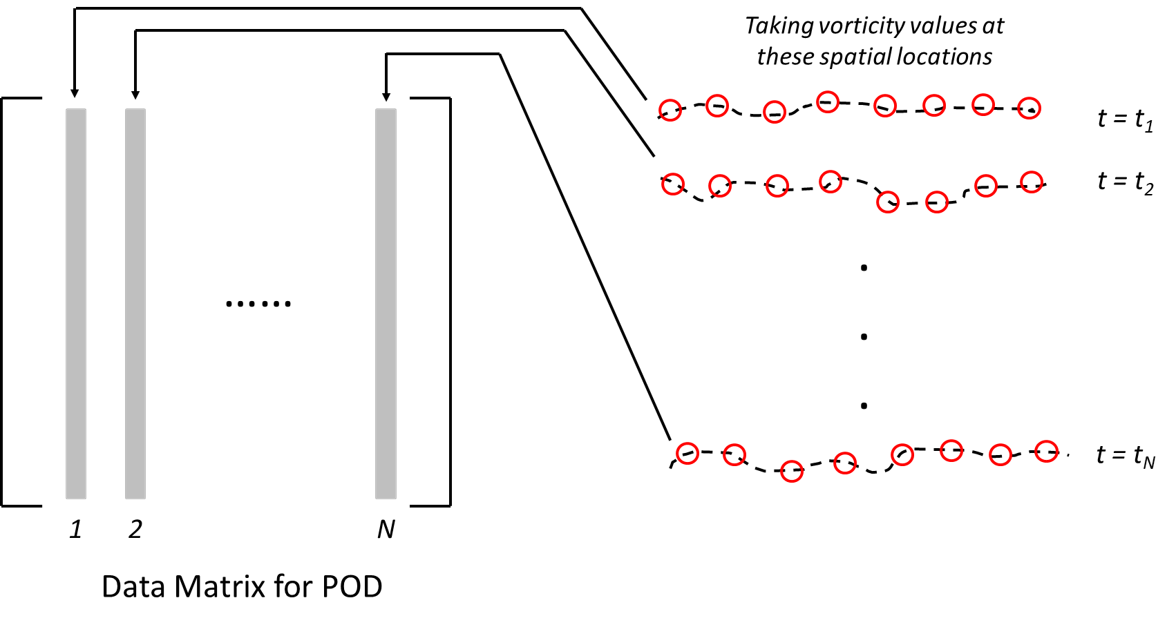}}
	\caption{Formulation of advective function using maximum vorticity}
	\label{f_Matrix}
\end{figure}

To explore the different possibilities of applying the proposed method to determine the convection velocity, the mean contour of a variable is shown in figure \ref{f_Mean_Vorticity} (vorticity is used here which is made non-dimensional using $\frac{r_0}{U_j}$). One approach is to populate a matrix with the data restricted to a sub-domain (indicated in the figure), at each instant of time and apply the technique. This strategy has been successfully demonstrated by \cite{Vincent_Jaunet} where the Schlieren image data of a high speed jet was used, the variable being the line-integrated gray-scale image intensity. However, an alternate approach can be used with a smaller spatial data size (1-D instead of 2-D). The important aspect for the application of the proposed method is the identification of an advection function along with the path of advection. In impinging jets, a significant role in the flow dynamics is played by the advection of the vortical structures in the shear layer. Hence, vorticity is believed to be the appropriate metric for the application of the technique. The mean radial location of the maximum vorticity at each streamwise location represents the path of advection in the shear layer, shown as dotted lines in figure \ref{f_Mean_Vorticity} and the corresponding value of the vorticity represents the advective function. The data set can now be created by populating the values of maximum vorticity at various instants of time, shown in figure \ref{f_Matrix}. If this data set is visualized as a video, one can indeed verify the process of advection.   

POD of the resultant data matrix gives the dominant modes. The most dominant mode, i.e. mode 1, represents the mean value and hence, plays no role in the analysis. The dot product of modes with their respective spatial derivatives as defined in equation \ref{e_test_con}, are found to have very low relative values for the second and the third modes. The convective velocity is therefore, computed using the combination of modes 2 and 3. The value determined using this methodology for the simulation data is 0.48 $U_j$ which is close to the estimation of 0.52 $U_j$ by \cite{Krothapalli}  for an experimental data set acquired at an order of magnitude higher Reynolds number. 

It is important to note that while the results are obtained using vorticity, a different metric can also be used if it shows advection along a certain path. For the present study, the fluctuating components of velocity were also tested but as expected, none of the two could alone capture advection on some fixed radial location.

\section{Sensitivity analysis on a model problem} \label{s_synthetic}

\begin{figure}
	\centerline{\includegraphics[scale=0.3]{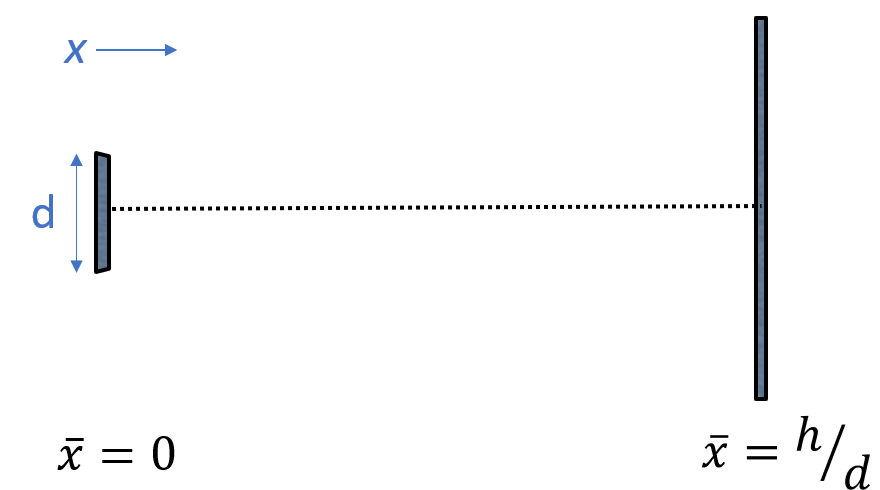}}
	\caption{Domain for the analysis of the model problem of impinging jet}
	\label{f_Model}
\end{figure}

An essential aspect to be examined is the sensitivity of the prediction with the time resolution of the data. A simple propagative wave model is used as the model problem for the aero-acoustic feedback loop whose domain is shown in figure \ref{f_Model}. The bar above the variables denotes the non-dimensional form of the variables. The spatial co-ordinate ($x$) and the impingement distance ($h$) are made non-dimensional with the nozzle exit diameter $d$. Defining the impingement tone frequency as $f$ and the corresponding time period as $T$, a synthetic 1-D data set for advection is generated by adding white noise to a sinusoidal traveling wave. The white noise is generated using a sequence of random numbers sampled from a standard normal distribution multiplied by an amplitude $A$. The advection function can be expressed as 
\begin{equation} \label{e_Synthetic}
f = \sin (kx - kU_c t) + A*Noise ,
\end{equation} 
where $U_c$ represents the convection velocity. Using the wave relation, 
\begin{equation} 
kU_c = 2 \pi f = \frac{2 \pi}{T} , 
\end{equation}
and the definition of the wavenumber,
\begin{equation}
k= \frac{2 \pi}{ \lambda } 
\end{equation}
where $\lambda$ represents the wavelength, the equation \ref{e_Synthetic} can be re-written as 
\begin{equation} \label{e_Synthetic_mod}
f = \sin \left[ 2 \pi \left( \frac{x}{\lambda} - \frac{t}{T} \right) \right] + A*Noise .
\end{equation} 
As the wavelength and the time period are the spatial and the temporal parameters of interest, respectively, it is prudent to non-dimensionalize the spatial variable with the wavelength and the temporal variable with the time period. Introducing $\tilde{x} = \frac{x}{\lambda}$ as another non-dimensional spatial variable and $\overline{t} = \frac{t}{T}$ as the non-dimensional form of time, the final form of the model equation for generating the synthetic data is  
\begin{equation} \label{e_Synthetic_final}
f = \sin \left[ 2 \pi \left( \tilde{x} - \overline{t} \right) \right] + A*Noise .
\end{equation} 
The convection velocity in this equation has a unit value in the non-dimensional form and hence, the proposed methodology when applied to this equation should ideally yield unity.

\begin{figure}
	\centerline{\includegraphics[scale=0.3]{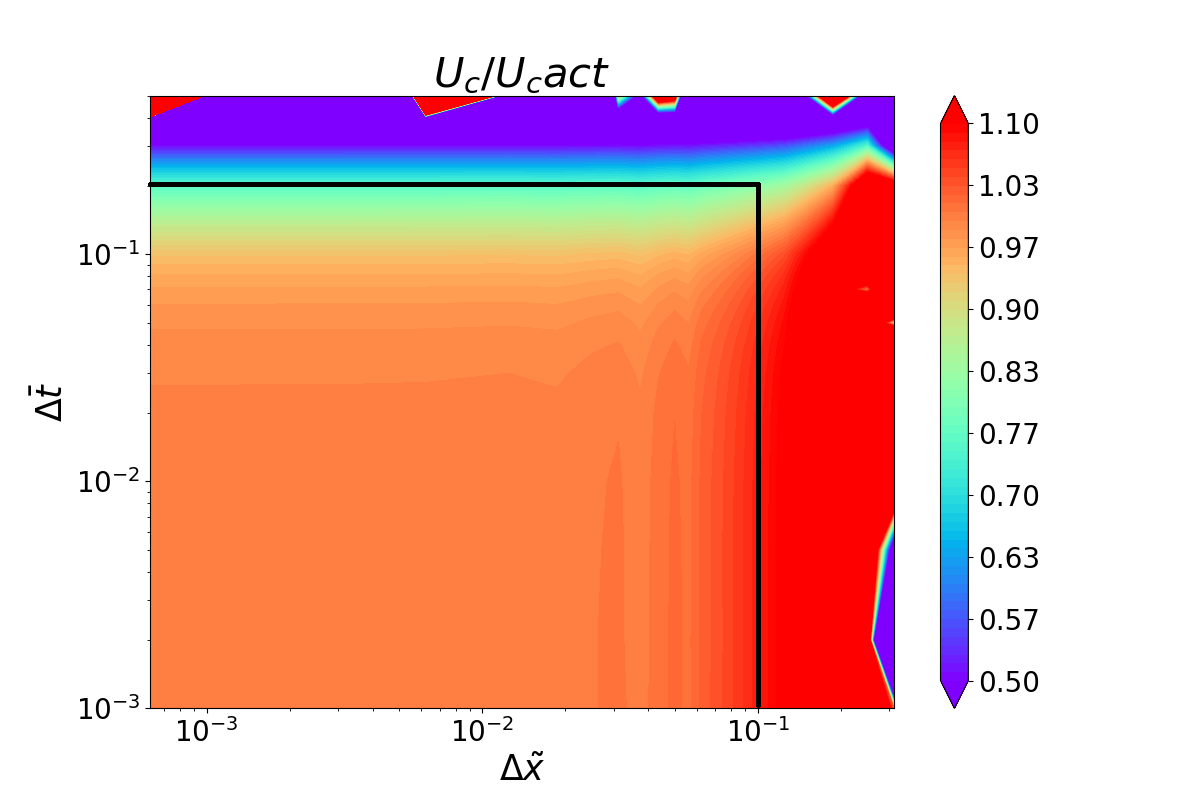}}
	\caption{Contours of the transfer function over a range of spatial and temporal resolutions }
	\label{f_Space_time_grid_contour}
\end{figure}

A high noise amplitude of 10 \% has been chosen for the analysis. As the synthetic data is generated using a well defined sinusoidal function, the first two modes will be the most dominant modes and also satisfy the condition for advection. The convection velocity is therefore, determined using the first two modes. Table \ref{t_grid} shows the range of grid resolutions over which the sensitivity analysis is performed. The ratio of the computed convection velocity to the actual one is used as a metric which can be termed as the transfer function, $\frac{U_c}{U_cact}$. The transfer function having unit value indicates that there is no bias in the predicted convection velocity, which represents the most ideal condition. Figure \ref{f_Space_time_grid_contour} shows the contours of the computed value over a domain of various temporal and spatial grid sizes. The reported values represent an average of 100 repetitions of the same computation. The uncertainty in the computed values was within 0.03 \%. As shown in the figure, the sub-domain inside the black lines has less than 20 \% bias in the computed values. This region corresponds to a sampling time of less than one-fifth of the phase cycle time period and a spatial resolution of less than one-tenth of the wavelength. However, for the bias to be within 1 \%, the sampling time should be of the order of one-hundredth of the phase cycle time period and the spatial resolution should be less than 20 times the wavelength ($\Delta \tilde{x} < \frac{\lambda}{20}$). 

\begin{table}
	\centering
	\begin{tabular}{|c|c|}
		\hline
		$\Delta \overline{t}$ & 0.001 - 0.5  \\ \hline
		$\Delta \tilde{x}$ & 0.0006 - 0.3 \\ \hline
		$N_t$ & 1000 \\ \hline
		$Noise$ & 10 \% \\ \hline
	\end{tabular}
	\caption{Computational parameters for the analysis}
	\label{t_grid}
\end{table}

\begin{figure}
	\centerline{\includegraphics[scale=0.3]{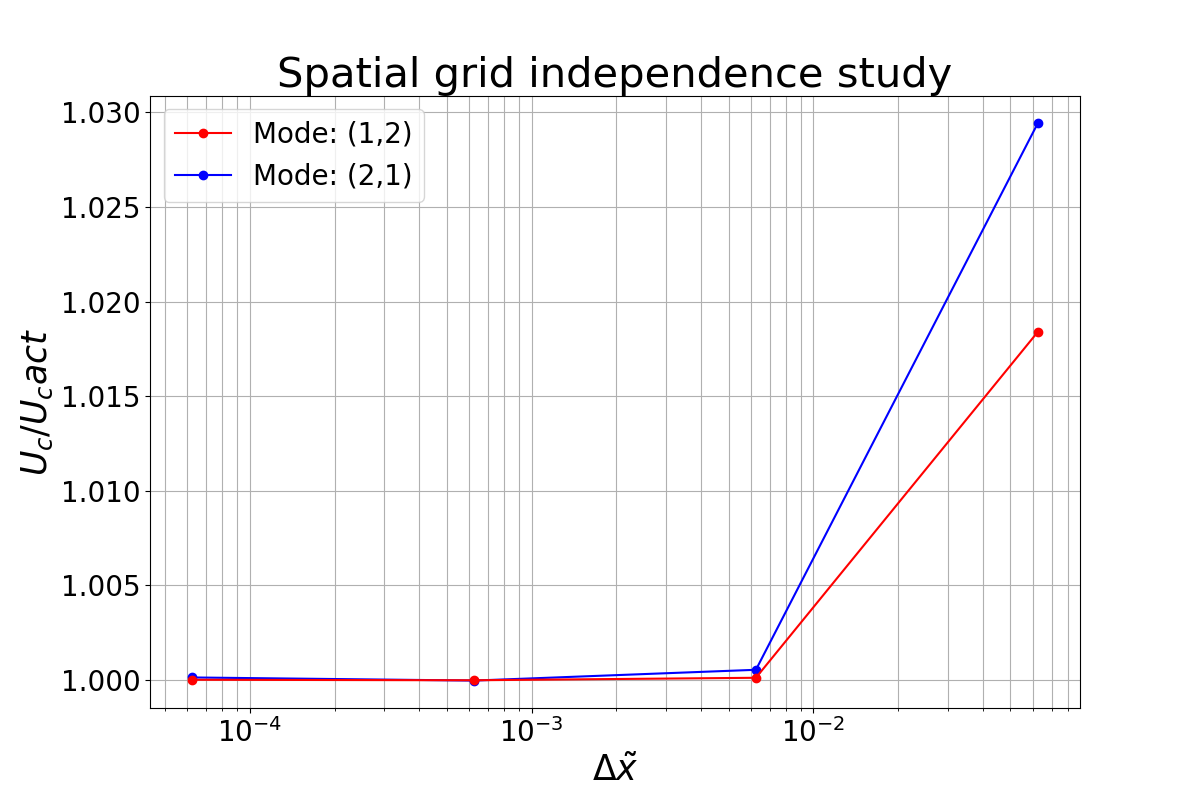}}
	\caption{Sensitivity of the transfer function to the spatial grid resolution for a time resolution of \textit{0.001 T}, over 10 phase cycles }
	\label{f_Spatial_Grid}
\end{figure}

\begin{figure}
	\centerline{\includegraphics[scale=0.3]{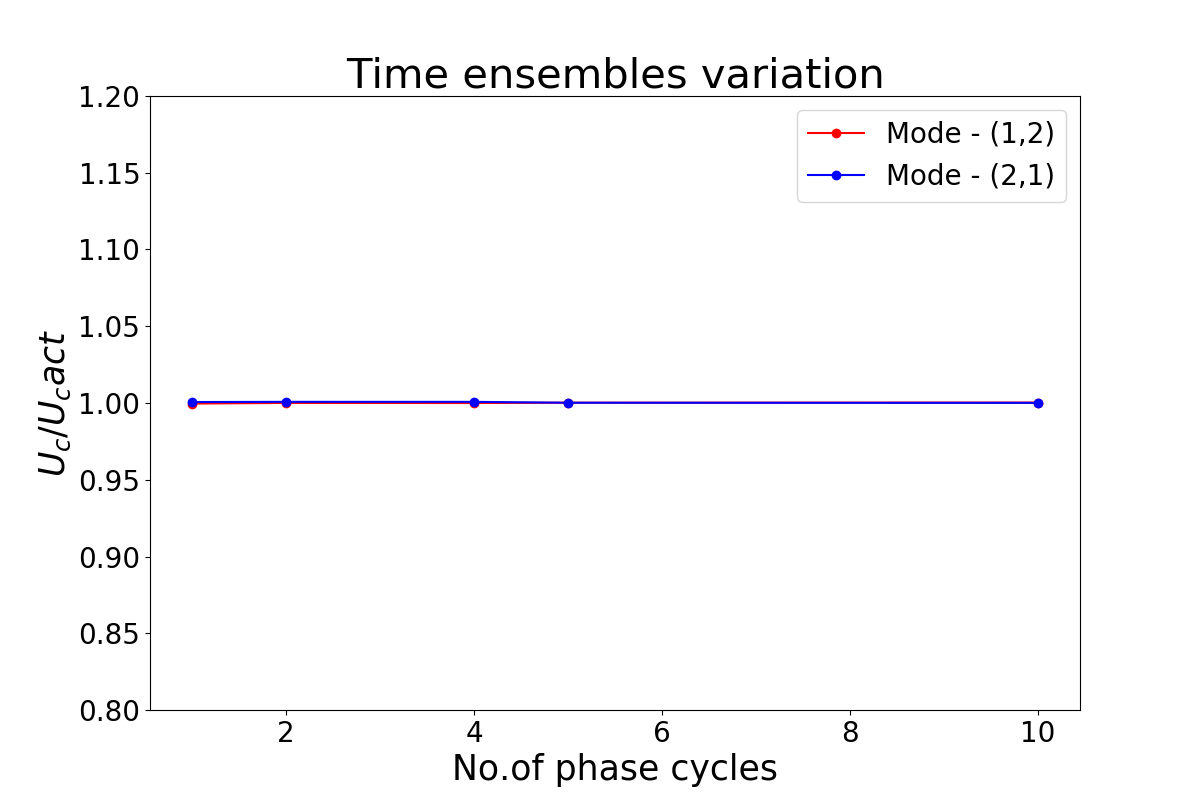}}
	\caption{Sampling size variation study for a time resolution of \textit{0.001 T} }
	\label{f_time_ensembles}
\end{figure}

Figure \ref{f_Spatial_Grid} shows the sensitivity with spatial grid resolution for a fixed time resolution of \textit{0.001 T}, over 10 phase cycles. It can be clearly observed that a spatial resolution of $0.006 \lambda$ is sufficient to capture the convection velocity.  As the data is time resolved, the time span of one feedback loop cycle is deemed sufficient. For the time resolution of \textit{0.001 T}, the number of required samples (\textit{$N_t$}) is 1000. However, as an additional sanity check, a temporal study is also shown in figure \ref{f_time_ensembles}. As expected, the predictions are found to be insensitive to the number of samples. 

\begin{table}
	\centering
	\begin{tabular}{|c|c|}
		\hline
		$A$ & 10 \%  \\ \hline
		$\Delta \tilde x$ & 0.006  \\ \hline
		$Nt$ & 1000 \\ \hline
	\end{tabular}
	\caption{Computational parameters for the analysis}
	\label{t_comp_par}
\end{table}

\begin{figure}
	\centerline{\includegraphics[scale=0.3]{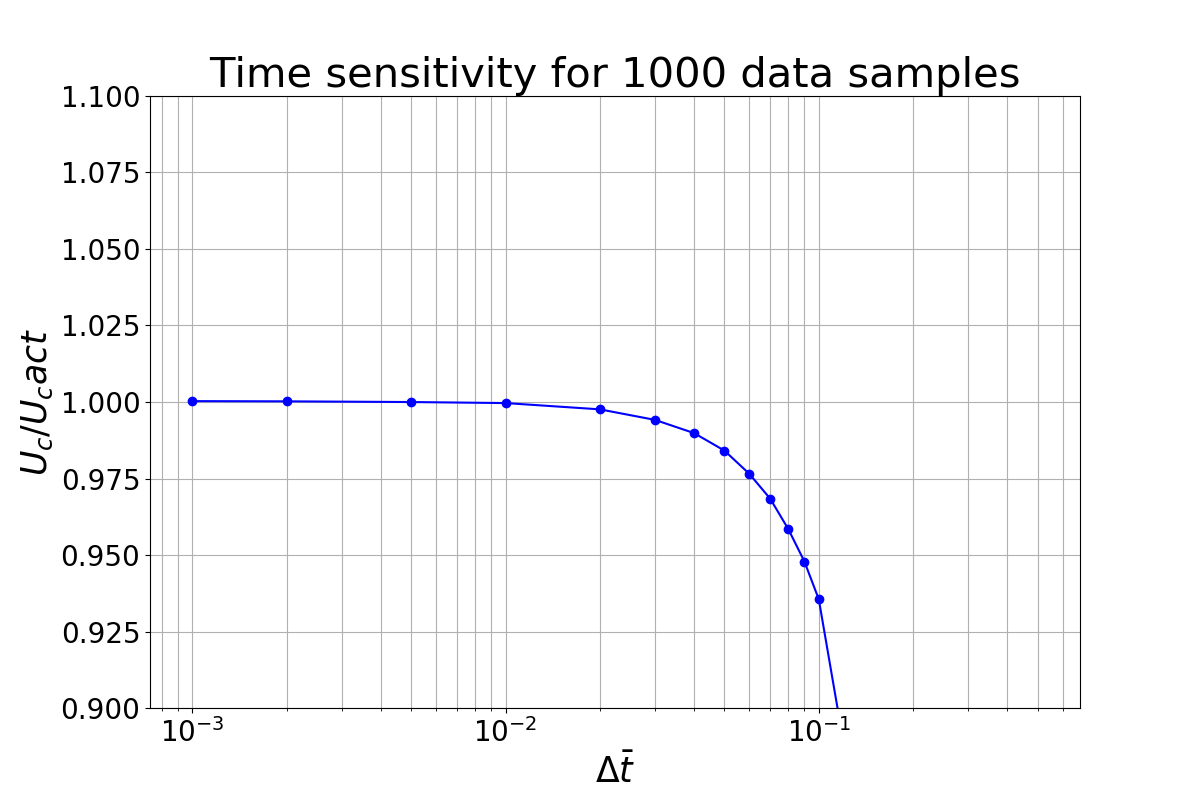}}
	\caption{Sensitivity of prediction with time resolution}
	\label{f_time_resolution}
\end{figure}

Based on the observations, table \ref{t_comp_par} summarizes the computational parameters for all further analysis. The computed values corresponding to the above parameters are shown in figure \ref{f_time_resolution}. The bias in prediction is 5\% for a time resolution of less than \textit{0.1 T}.  For a practical interpretation, especially for experimental research, the impingement tones can be considered to be of the order of 10 kHz (\cite{Krothapalli} and \cite{Joel}). This implies a time period of \SI{100}{\micro s} and hence, a time resolution of at least \SI{10}{\micro s} for a reasonable estimation and about \SI{1}{\micro s} for prediction having a very low bias.

For the research studies in high speed impinging jets, the wavelengths are often expressed in terms of the nozzle exit diameter ($d$) as shown in \cite{Tam_Review}, \cite{Sinibaldi} and \cite{henderson2005}. The spatial parameters in the experimental studies using imaging techniques are also specified in terms of the diameter (\cite{Joel}, \cite{Joel_JFM_2} and \cite{HS_Imaging_3}). Hence, conversion from one form of the non-dimensional spatial variable to the other can be useful for the research applications. Using the definition of Strouhal number,
\begin{equation}
St = \frac{fd}{U_j} ,
\end{equation}
and defining 
\begin{equation}
\overline{U_c} = \frac{U_c}{U_j} 
\end{equation}
as the scaled convection velocity, the two non-dimensional spatial variables can be related as

\begin{equation} \label{e_convert}
\overline{x} = \tilde{x} \left( \frac{\lambda}{d} \right) = \tilde{x} \left( \frac{\overline{U_c}}{St} \right) .
\end{equation}   
The conversion factor in equation \ref{e_convert} depends on the flow dynamics of the particular research problem. 

In context of the aero-acoustic feedback loop, it is important to note that the flow parameters ($\overline{U_c}$ and $St$) are not independent. They are related by Powell's equation for a feedback loop (\cite{powell}), which can be expressed in non-dimensional form as
\begin{equation} \label{e_Powell_ND}
\frac{N+p}{St} = \overline{h} \left( M_j + \frac{1}{\overline{U_c}} \right ) .
\end{equation}
To evaluate the spatial scales in terms of the diameter ($d$), some results from the literature were used as the sample data points. The experiments by \cite{Krothapalli} using a converging-diverging (C-D) nozzle operating at an NPR (nozzle pressure ratio) of 3.7 showed an estimate of 0.52 for the non-dimensional convection velocity with the Strouhal number measured to be 0.324. For an additional sanity check, the parameter $N$ was calculated using a phase lag (p) of -0.49 in equation \ref{e_Powell_ND}, as proposed by \cite{Joel}. It turns out to be an integer with value 5 which shows compliance with the Powell's equation.  Another sample data was taken from one of the experiments by Sinibaldi et al. (\cite{Sinibaldi}) using a choked nozzle operated at $NPR =2.05$ which yielded $St = 0.492$ and $\overline{U_c} = 0.58$. The parameters for both the configurations are summarized in table \ref{t_test_mat} where $N_{\lambda}$ refers to the number of wavelengths in the impingement distance ($\bar{h}$). Using the equation \ref{e_convert}, the required minimum spatial resolution for low bias ($\Delta \tilde{x} = \frac{\lambda}{20}$) turns out to be $0.1d$ for the first configuration. However, for the second configuration, the same resolution corresponds to a spatial scale of $0.06d$. Hence, it is important to consider the wavelengths expected in the flow dynamics for each experiment or simulation as the spatial resolution which is adequate for a particular operating condition might not be sufficient at a different flow condition.  

\begin{table}[ht] 
	\centering
	\begin{tabular}{|c|c|c|}
		\hline
		\textit{Parameter}  & \textit{Configuration 1} & \textit{Configuration 2}  \\ 
		$\overline{h}$ & 4 & 4  \\ 
		$\overline{U_c}$ & 0.52 & 0.58  \\
		$St$ & 0.324 & 0.492 \\ 
		$\frac{\lambda}{d}$ & 1.6 & 1.2 \\
		$N_{\lambda}$ & 2.5 & 3.3 \\ \hline
	\end{tabular}
	\caption{Some sample parameters obtained from the literature data}
	\label{t_test_mat}
\end{table}

\section{Application to a data set which is not time resolved} \label{s_Double_PIV}
\begin{figure}
	\centerline{\includegraphics[scale=0.3]{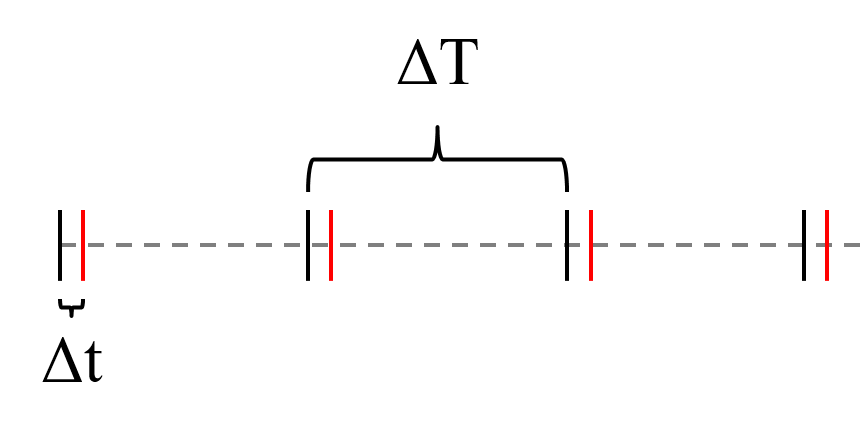}}
	\caption{Experimental data to be tested for the application of the proposed method}
	\label{f_exp_data}
\end{figure}

As shown in section \ref{s_synthetic}, the time resolved velocity measurements using PIV techniques requires a high time resolution of about \SI{1}{\micro s}, which is challenging where high spatial resolution is also required. The question, therefore, arises whether the entire data set needs to be time resolved for the application of the proposed methodology. Closer examination of the underlying principles behind the technique indicates that the data might not necessarily be time resolved for the entire time, and that time resolution in the data is only required for the computation of the instantaneous temporal derivatives. If it is possible to obtain these derivatives by an alternate approach, the technique can be applied to experimental data obtained using high spatial resolution PIV. An example of the data set required is shown in figure \ref{f_exp_data} which is representative of a data set acquired using two acquisition systems that allows for the measurement of two fields separated by a small time $\Delta t$. The data from one system is shown in black and the data from the other is shown in red. Such data set was experimentally generated in 1989 by \cite{papamoschou1989two} for high speed flows using two-spark schlieren system. The convection velocity was estimated from the images generated using two-spark schlieren system in 2016 by \cite{Vincent_Jaunet} with the grayscale image intensity as the data input. Similar data set for velocity field was obtained using dual-time stereoscopic particle image velocimetry (PIV) by \cite{perret2006}. As shown by \cite{Jonathan_DMD}, such data sets can be used for the determination of the instantaneous temporal growth rates using DMD (dynamic mode decomposition). Moreover, as the time resolution can be specified directly by the time offset in the data captured by the two camera systems, one can obtain high temporal resolution of the order of \SI{100}{ns}, which cannot be obtained by high speed camera. Such data set shall be referred to as the double PIV data for all further references.

\begin{figure} 
	\begin{subfigure}{0.5\textwidth}
		\centering
		\includegraphics[height=1.25in]{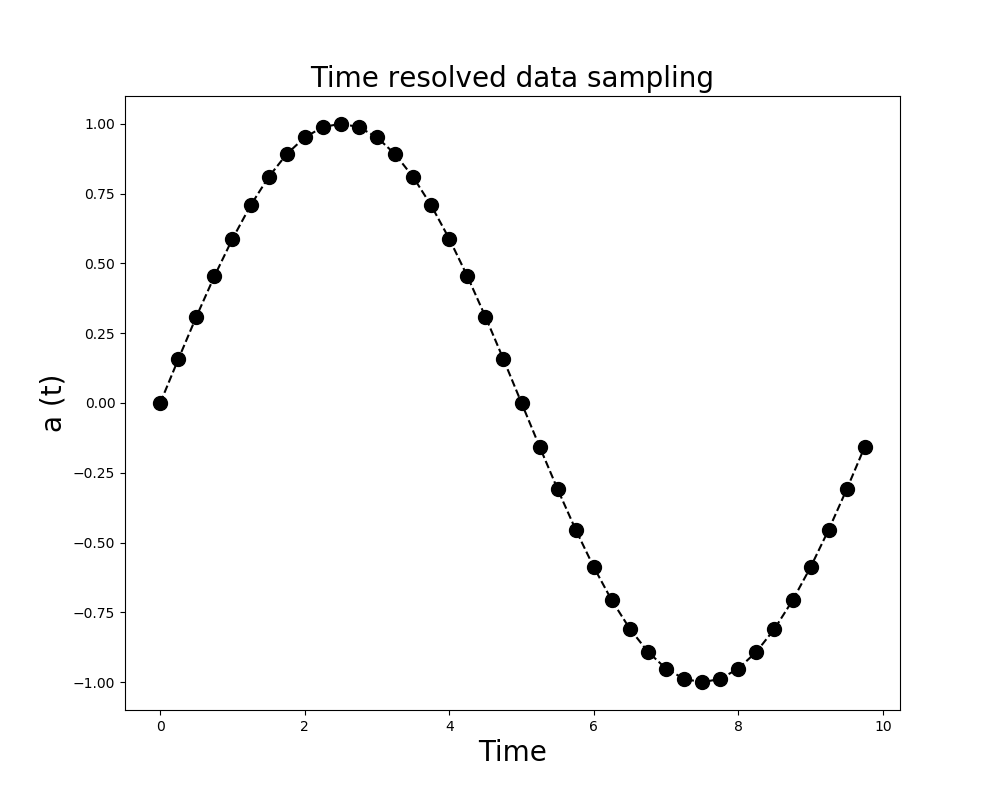}
		\caption{}
	\end{subfigure}
	\begin{subfigure}{0.5\textwidth}
		\centering
		\includegraphics[height=1.25in]{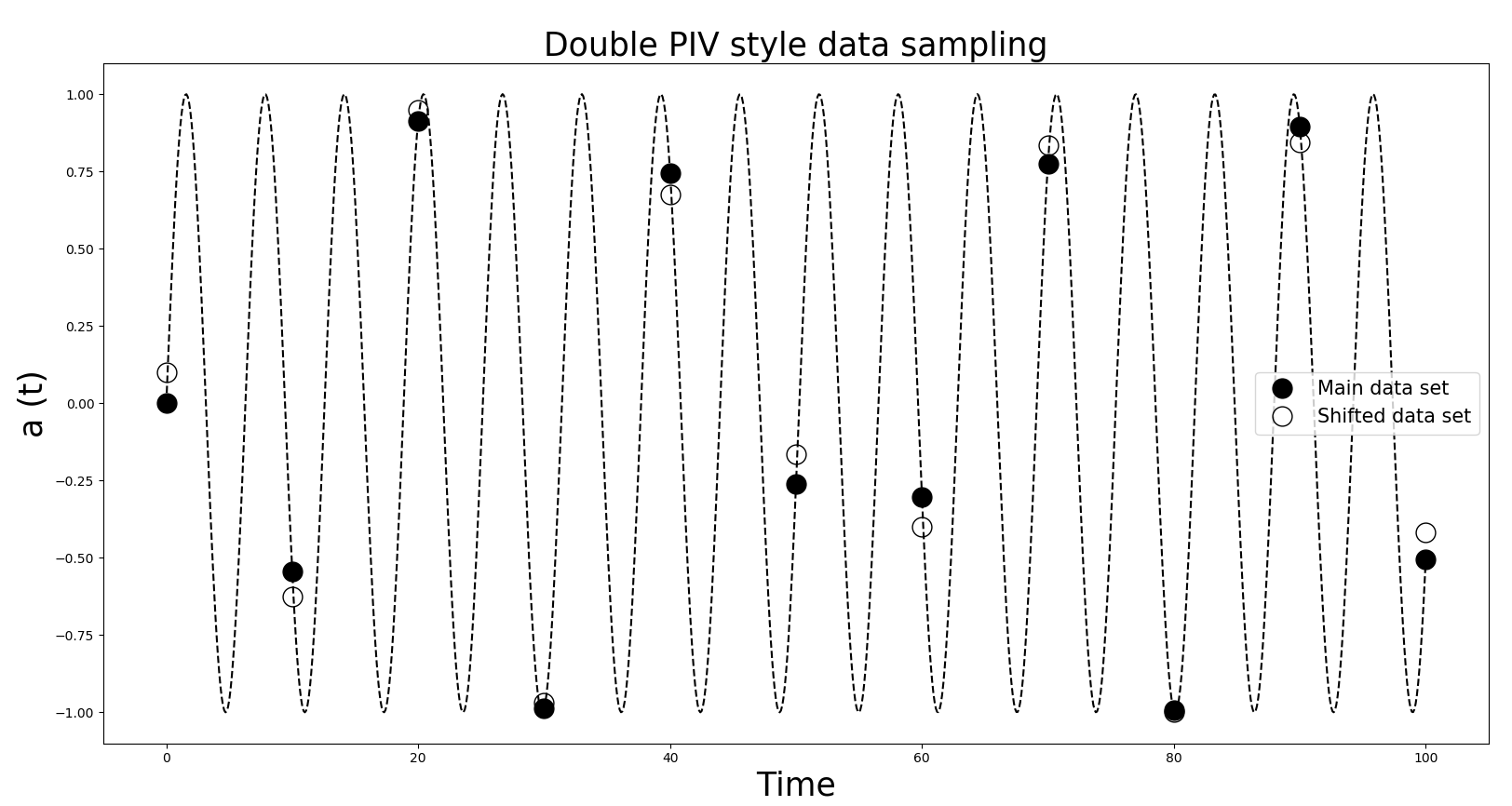}
		\caption{}
	\end{subfigure}
	\caption{Plots of sinusoidal functions showing (a) Time resolved data and (b) Double PIV style data}
	\label{f_POD_analysis}
\end{figure}

\begin{figure}
	\centerline{\includegraphics[scale=0.3]{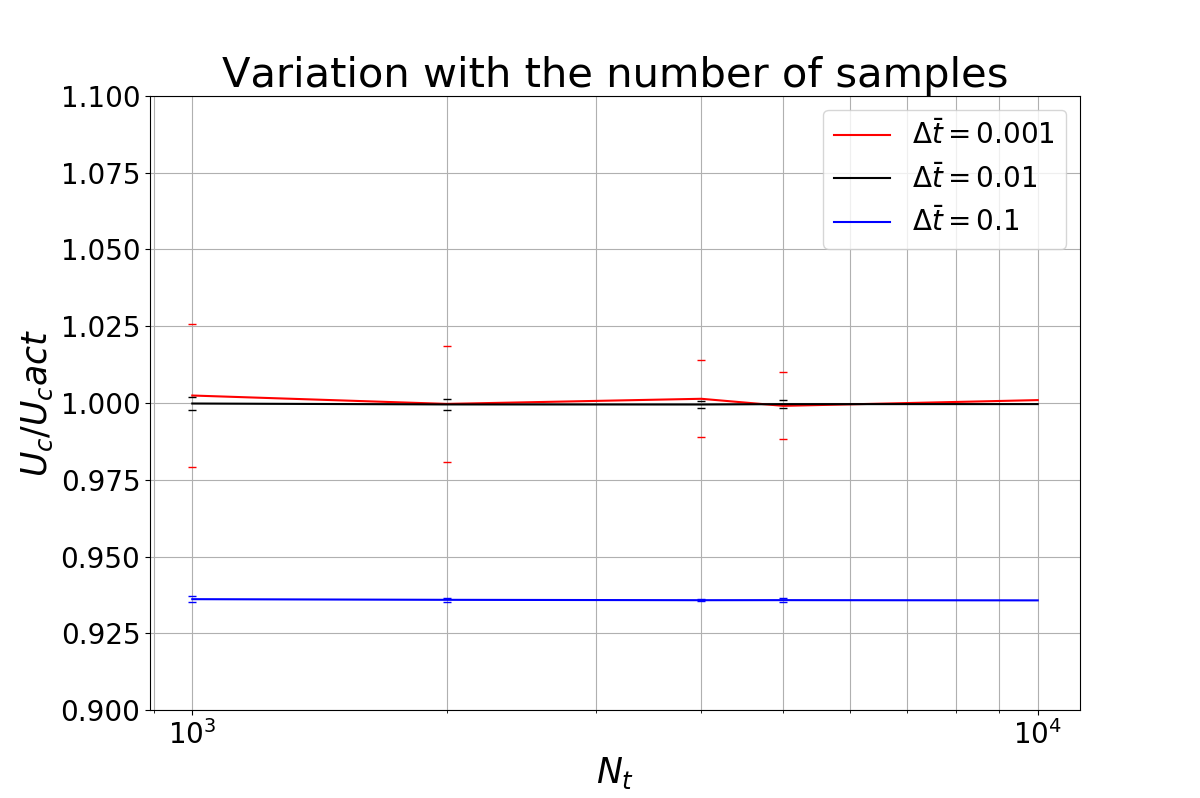}}
	\caption{Sensitivity study with the number of samples for double PIV style sampled data}
	\label{f_Sample_study}
\end{figure}

This methodology was tested using a synthetic double PIV data set generated using the same function as that used in section \ref{s_synthetic}. Since the two data sets in this case are not time resolved, $\Delta T$ is orders of magnitude larger than the time shift $\delta t$ and hence, there is no correlation between two successive samples of any two data sets as shown in figure \ref{f_POD_analysis} (b). Successive samples of the main data set were therefore, generated using a random number generator. The noise level and the spatial resolution was kept the same as that in table \ref{t_comp_par}. A sensitivity study with the number of samples was carried out whose results are shown in figure  \ref{f_Sample_study}. The uncertainty was within 3\% for all the cases with the maximum value for the smallest time resolution, $ \overline{\Delta t} = 0.001 $. As illustrated in \cite{Tushar}, a small time resolution in the presence of significant noise can result in larger deviation from the expected value due to the possibility of some erroneous instantaneous time derivatives. However, as the maximum uncertainty is still very low, 1000 samples are deemed sufficient for any further analysis. 

\begin{figure}
	\centerline{\includegraphics[scale=0.3]{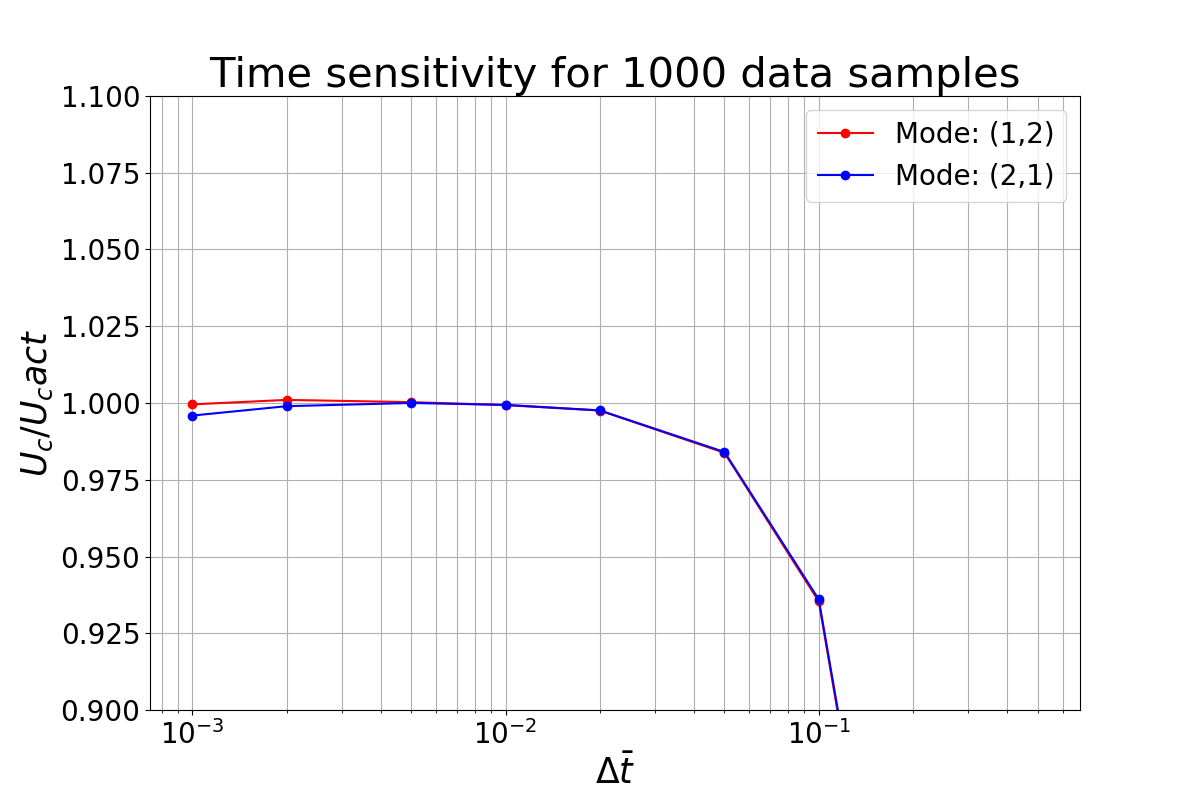}}
	\caption{Time sensitivity study for double PIV style sampled data}
	\label{f_time_sensitivity_Double_PIV}
\end{figure}

Figure \ref{f_time_sensitivity_Double_PIV} shows the sensitivity of the prediction of the convection velocity with time resolution. The trend is identical to that observed in figure \ref{f_time_resolution}. Hence, whether the data is time resolved or sampled like a double PIV data, the requirement of the time resolution for obtaining low bias in the prediction, stays the same. This implies that the experimental investigation of the flow dynamics of a supersonic jet can be undertaken without a time resolved time series data. Measurements obtained using double PIV can provide the required dynamical information if an adequate time shift is specified between the two data sets.

\section{Conclusion}
The POD-Galerkin based advection model can provide a good estimate of the convective velocity of the large scale coherent structures in the jet shear layer in supersonic free and impinging jets. The technique is relatively simple to implement and computationally less expensive than some other existing methodologies. A valuable advantage of the methodology is that it does not require time resolved time series data, provided the instantaneous temporal derivatives can be computed. Though the methodology is motivated by supersonic impinging jets, its application goes well beyond into other supersonic flows, combustion or heat transfer related problems.

As shown in the previous section, the technique eliminates the necessity of using high speed time series imaging techniques for the experimental study of supersonic jets. Data sets obtained using two PIV systems, which are offset by a small time $\Delta t$, are sufficient to capture the convection velocity of the large scale coherent structures in the shear layer. Though the prediction using this methodology are dependent on the spatial, as well as the temporal resolution, the sensitivity is very low over a broad range of grid resolutions. For experimental studies employing imaging techniques, the spatial resolution is generally governed by the camera sensor while the temporal resolution is controllable or selectable to a limit governed by the bandwidth of the hardware. For a fixed spatial resolution, a time resolution of less than one-tenth of the phase cycle time period is desirable to capture the convection velocity. 

\section{Acknowledgements}
The research is being funded by a Discovery Project Grant from the Australian Research Council (ARC), which is gratefully acknowledged.

\bibliography{mybibfile}

\end{document}